# Practical Simplified Indoor Multiwall Path-Loss Model

Taewon Kang and Jiwon Seo[*]

School of Integrated Technology, Yonsei University,
Incheon, 21983, Korea (taewon.kang, jiwon.seo@yonsei.ac.kr)
Incheon, 21983, Korea
* Corresponding author

**Abstract**: Over the past few decades, attempts had been made to build a suitable channel prediction model to optimize radio transmission systems. It is particularly essential to predict the path loss due to the blockage of the signal, in indoor radio system applications. This paper proposed a multiwall path-loss propagation model for an indoor environment, operating at a transmission frequency of 2.45 GHz in the industrial, scientific, and medical (ISM) radio band. The effects of the number of the walls to be traversed along the radio propagation path are considered in the model. To propose the model, the previous works on well-known indoor path loss models are discussed. Then, the path loss produced by the intervening walls in the propagation path is measured, and the terms representing the loss factors in the theoretical path-loss model are modified. The analyzed results of the path loss factors acquired at 2.45 GHz are presented. The proposed path-loss model simplifies the loss factor term with an admissible assumption of the indoor environment and predicts the path-loss factor accurately.

**Keywords:** Multiwall model, path loss, ISM radio band

## 1. INTRODUCTION

The increased demands of indoor wireless communication applications have drawn much interest in the study of indoor radio propagation. With the increase in the use of radio systems, including global navigation satellite systems [1-6] and long-range navigation (Loran) [7-9], many applications with various purposes have been developed [10-16]. Consequently, an accurate channel prediction model for wideband channels has become essential to researchers. As signal reception environments affect the radio systems and system applications greatly [17-19], it is important to predict the path loss due to the signal blockage, reflection, or interference [20]. Especially, in indoor environments, signal blockage due to walls and floors affects radio transmission and reception, greatly, exhibiting rapid changes in the path loss for each location [21, 22].

To obtain the path loss factors due to signal blockage in indoor environments, RF transmission and reception experiments are necessary. By receiving RF signals from multiple positions with different signal blockage factors, the path loss model can be obtained with less prediction errors.

Various indoor-channel path-loss models have been developed in the past [23-25]. Although some of these models can be used to predict the path loss in certain indoor environments for a wide range of frequency bands, prediction errors might occur at other frequencies or for other indoor materials [26].

In this paper, a path loss model that considers multiple intervening walls is proposed. With an admissible assumption of the indoor environment, the loss factor term was simplified. The indoor path loss models from the previous works are discussed in Section 2. The experimental results and a description of the proposed model, in accordance with the results, are presented in Section 3. The conclusions are presented in Section 4.

## 2. INDOOR PATH-LOSS MODEL

One of the basic path-loss models for the indoor radio channel is the one-slope model [27], where the free-space path-loss term is introduced.

$$PL_{one-slope} = PL_0 + 10n \log(d), \quad (1)$$

where $PL_{one-slope}$ is the path loss in dB, $PL_0$ is the reference path loss, which is the path loss over a distance of one meter, $d$ is the distance between the transmitter and receiver, and $n$ is the path-loss exponent that indicates how fast the path loss increases with distance.

Previous research on radio propagation in indoor environments [28] represented all penetrated walls, using individual penetration losses depending on their category, which was classified in terms of the thickness and material. Walls within the same category were found to contribute a constant loss, irrespective of whether other walls or floors had been penetrated before.

In accordance with this information, the COST231 multiwall model is represented as follows [29].

$$PL_{cost} = PL_0 + 10n \log(d) + \sum_{i=1}^{M} PL_i, \quad (2)$$

where $M$ is the total number of walls to be traversed in the path along which the radio signal propagates and $PL_i$

denotes the path loss due to $i$th intervening wall.

## 3. PROPOSED MULTIWALL PATH-LOSS MODEL

### 3.1 Proposed indoor multiwall path-loss model

The path loss terms introduced by multiple walls in the previous multiwall model [29] described the path loss created by the individual walls, using multiple separate values. Our indoor multiwall path-loss model assumes the case wherein all the walls in the indoor environment belong to a single category, which can be applied in the case of a single building floor with identical walls. In this case, the term representing the summation of the path losses created by the multiple intervening walls can be rewritten in terms of the path loss of single intervening wall multiplied by the total number of intervening walls. Thus, this practical model requires less experimental path loss values of intervening wall.

$$PL = PL_0 + 10n\log(d) + M \times PL_w, \quad (3)$$

where $M$ is the total number of intervening walls and $PL_w$ denotes the path loss of a single traversing wall.

### 3.2 Experimental results

To verify the path loss of an intervening wall, we conducted signal transmission and reception experiments on the first floor of the Veritas Hall C building (Yonsei University, Incheon, South Korea). Fig. 1 shows the floor plan of the test site. The transmission was set up using a HP ESG-D300A digital RF signal generator as the transmitter, and was received using an Anritsu MS2712E Spectrum Master as the receiver. The radio signal was transmitted at a center frequency of 2.45 GHz with 100 MHz bandwidth, which was the radio spectrum reserved internationally for industrial, scientific, and medical (ISM) purposes, and could be transmitted and received locally for scientific research [30]. The experiment was conducted with all the doors on the floor closed. The floor plan of the building was used to obtain the straight-line distance between the transmitter and receiver. The intervening walls were constructed of cement mortar and had a thickness of 25 cm.

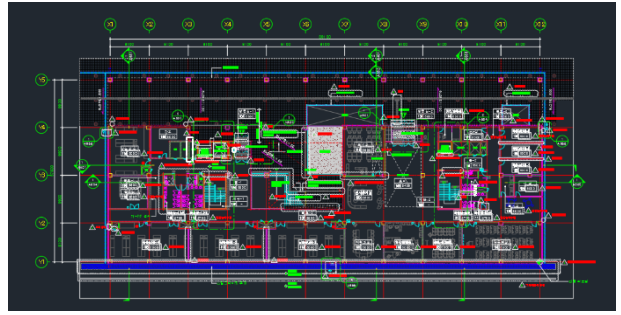

Fig. 1 Architectural floor plan of the test site (building)

The transmission power was set to 20 dBm. The path loss introduced by the intervening walls was calculated by comparing the received signal strength at a straight-line distance, with and without intervening walls. Table 1 shows the experimental results of the path loss caused by the intervening walls, measured with respect to the number of intervening walls along the propagation path. Fifty measurements were taken in each case.

Table 1 Experimental results of the path loss introduced by intervening walls.

| No. of intervening walls | Path loss by wall (dB) | Standard deviation (dB) | Straight-line distance (m) |
| --- | --- | --- | --- |
| 1 | 18.62 | 1.26 | 2.15 |
| 2 | 35.86 | 2.92 | 3.95 |
| 3 | 52.87 | 2.96 | 6.5 |

The experiment showed that the path loss introduced by the intervening walls belonging to a single category showed a linear increase when the number of intervening walls increased. By fitting the path loss value linearly with the intervening wall number, the empirical value for $PL_w$ of the cement mortar wall with 25 cm thickness was obtained as 17.78 dB/wall, from the experiment. The experimental value of the path-loss exponent, $n$ in (3), was obtained as 3. Fig. 2 shows the path loss values measured for multiple intervening walls and the obtained path-loss value per wall.

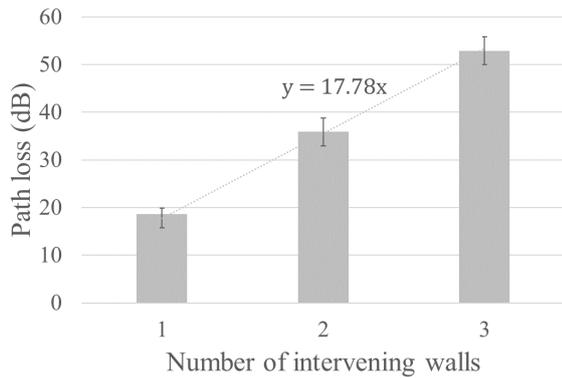

Fig. 2 Empirical values of the path losses produced by multiple intervening walls and the path loss per wall.

## 4. CONCLUSION

This paper presented a path-loss radio propagation channel model for indoor environments. A practical multiwall model was developed, which denoted the path loss introduced by multiple intervening walls along the radio propagation path by a multiplication of the path loss of a single intervening wall with the total number of intervening walls. The proposed model requires less experimental path loss values of intervening wall, with an admissible assumption of the indoor environment. This model can be applied in a single-floor indoor environment where the thickness and material of all walls are identical. Experiments were conducted in the 2.45 GHz ISM band in a single-floor indoor environment and a path-loss factor for a single intervening wall was obtained from the experimental results. Because of the simple structure of the proposed path-loss model, it could be applied to indoor radio-propagation applications such as ray tracing or indoor localization, with ease.

## ACKNOWLEDGEMENT

This work was supported by Institute for Information & Communications Technology Planning & Evaluation (IITP) grant funded by the Korea government (KNPA) (No. 2019-0-01291, LTE-based accurate positioning technique for emergency rescue).